\documentclass[12pt]{iopart}
\usepackage{iopams}
\usepackage[T1]{fontenc}
\usepackage[cp1251]{inputenc}
\usepackage{longtable,supertabular}

\def\v1{\vspace{1cm}}
\def\be{\begin{equation}}
\def\ee{\end{equation}}
\def\bc{\begin{center}}
\def\ec{\end{center}}

\def\vh{\varphi}

\newcommand{\bea}{\begin{eqnarray}}
\newcommand{\eea}{\end{eqnarray}}

\begin{document}

\title{Hamiltonian Approach to Conformal Coupling\\
 Scalar Field in the General Relativity}

\author{\L ukasz A. Glinka}
\address{Bogoliubov Laboratory of Theoretical Physics,\\ Joint Institute for Nuclear Research,\\ 6 Joliot--Curie Street, 141980
Dubna, Russia} \ead{glinka@theor.jinr.ru}

\author{Victor N. Pervushin}
\address{Bogoliubov Laboratory of Theoretical Physics,\\ Joint Institute for Nuclear Research,\\ 6 Joliot--Curie Street, 141980
Dubna, Russia} \ead{pervush@theor.jinr.ru}

\author{Ryszard P. Kostecki}
\address{Chair of Theory of Relativity and Gravitation,\\ Institute of Theoretical Physics, Warsaw University,\\ ul. Ho\.za 69, 00-681, Warsaw, Poland}
\ead{Ryszard.Kostecki@fuw.edu.pl}

\begin{abstract}

 \indent The dynamic status of scalar fields is studied in  the Hamiltonian approach to the General Relativity. We
show that the conformal coupling of the scalar field violates the
standard geometrical structure of the Einstein equations in GR and
their solutions including the Schwarzschild one and  the Newton
static interaction.

 In order to restore the  standard geometrical structure of GR, the scalar field is mixed with the scale metric component by the
Bekenstein type transformation. This ``scalar-scale'' mixing
converts the conformal coupling scalar field with  conformal weight
(n= --1) into the minimal coupling scalar field with zero conformal
weight (n=0) called a "scalar graviton".

Cosmological consequences of the "scalar-scale" mixing are
considered in the finite space-time  by extraction of the zero mode
(homogeneous) harmonics of a "scalar graviton". The classical
dynamics of "scalar graviton" testifies about a tremendous
contribution of its kinetic energy into the Universe evolution at
the beginning in the form of the rigid state.
\end{abstract}


\maketitle

\section{Introduction}

 The scalar fields play very important role in both
   particle physics  as the Higgs field in Standard
 Model (SM) and  standard  cosmological model  as
 one of elements  of the inflation mechanism \cite{linde}.
 A scalar field (SF) can be  introduced into
  General Relativity (GR) in two manners:
  without $R/6$ term (a minimal coupling)
  and with $R/6$ term (a conformal coupling) \cite{pct}.
  The difference between two couplings becomes essential
  at cosmological applications and the unification of SM
  with GR. Therefore, the research  of difference
  of these two couplings from the
  dynamic point of view is the topical problem \cite{PPG,PS}.

 The present paper is devoted to investigation of
 dynamical status of different
  couplings of scalar field by means of the Hamiltonian approach to GR
  \cite{dir,ADM,WDW,M,ps1}
  generalized for  finite space in \cite{242,242a} and
   the Lichnerowicz transformation
   to  the unit  determinant
   of the spatial metric \cite{lich}. We research
  self-consistences of initial conditions with equations of motion
   and boundary conditions, with the variation problem.

   In Section 2, the status of a conformal  coupling scalar field in GR is
    considered. In Section 3, the correspondence of the considered model
    with a relativistic brane is established.
     Sections 4, 5 are devoted to the Hamiltonian approach
    to the considered model in the finite space-time. In Section 6,
 cosmological consequences  are studied.

\section{Scalar field: action, interval, and symmetries}

 The sum of GR and  SF
 actions is
 \be\label{21-1}
 S_{GR+SF,\xi}=\int d^4x\sqrt{-g}
 \left[-\frac{\vh_0^2}{6}R(g)+\xi\frac{\Phi^2}{6}R(g)
 +g^{\mu\nu}\partial_\mu\Phi\partial_\nu\Phi
 \right],
 \ee
 where
 \be\label{1-h2}
 \vh_0=\sqrt{\frac{3}{8\pi}}\mathcal{M}_{Pl}\simeq 0.3 \cdot 10^{18} {\rm GeV}
 \ee
 is the mass scale factor,
 $R(g)$ is the curvature Ricci scalar and $g_{\mu\nu}$ is the metric
 tensor on the Riemann manifold with the {\it``geometric interval''}
 \be\label{1-h7s}
 {ds^2} = g_{\mu\nu}dx^\mu x^\nu,  \ee
$\xi=0,1$ for minimal ($\xi=0$) and conformal ($\xi=1$) couplings.
 The {\it``geometric interval''} can be written in terms of
 linear differential forms as components of an orthogonal
 Fock's simplex of reference $\omega_{(\alpha)}$
 \be \label{ds}
ds^2\equiv\omega_{(\alpha)}\omega_{(\alpha)}=
 \omega_{(0)}\omega_{(0)}-
 \omega_{(1)}\omega_{(1)}-\omega_{(2)}\omega_{(2)}-\omega_{(3)}\omega_{(3)}.
 \ee

 In terms of  simplex the GR
  contains  two principles of
 relativity: the {\it``geometric''} in the form of general coordinate transformations
\bea \label{1zel}
 x^{\mu} &\to&  \tilde x^{\mu}=\tilde
 x^{\mu}(x^0,x^{1},x^{2},x^{3})\\
 \omega_{(\alpha)}(x^{\mu})&\to&\omega_{(\alpha)}(\tilde x^{\mu})=
 \omega_{(\alpha)}(x^{\mu})
 \eea
 and the {\it``dynamic''} principle formulated as the Lorentz
 transformations of an orthogonal  simplex of reference
 \be \label{2zel}
{\omega}_{(\alpha)}~\to ~
\overline{\omega}_{(\alpha)}=L_{(\alpha)(\beta)}{\omega}_{(\beta)}.
\ee
  The latter are considered as transformations of a frame of reference.

\section{Conformal coupling scalar field in GR as a
relativistic brane}

\indent The conformal coupling scalar field action
 \be\label{2sc}
 S_{SF,\xi=1}=\int d^4x
 \left[\sqrt{-g}\frac{\Phi^2}{6}R(g)
 -\Phi \partial_\mu (\sqrt{-g} g^{\mu\nu}\partial_\nu\Phi)
 \right],
 \ee
 is invariant with respect to
  scale transformations of metric components and scalar
  field
 \be\label{ct}
 g_{\mu\nu}^{\Omega}=\Omega^{2}
 g_{\mu\nu},~~\Phi^{\Omega}=\Omega^{-1}\Phi
 \ee
 with the conformal
 weights  $n=2,-1$ for the tensor and scalar
 fields, respectively.
 The conformal coupling of the scalar field
 with the weight $n=-1$ is required by
 unification of GR and
SM. The latter is scale invariant
 except of the Higgs potential. However, the Hilbert -- Einstein
 action $S_{GR}$ in Eq. (\ref{21-1}) is not invariant.
 After the scale transformation
 the total action (\ref{21-1}) $S_{GR+SF,\xi=1}$
 takes the form of the conformal relativistic brane
 \bea\label{brane-m}
&&S_{GR}[g^{\Omega}]+S_{SF,\xi=1}[g^{\Omega},\Phi^{\Omega}]=
S_{\mathrm{brane}}^{(D=4/N=2)}[X_{(0)},X_{(1)}]\!=\nonumber\\&-&\!\int\!d^4x\!\Bigg[\sqrt{-g}\!\frac{X_{(0)}^2-X_{(1)}^2}{6}\,{}^{(4)}\!R(g)-X_{(0)}\partial_\mu(\sqrt{-g}g^{\mu\nu}\partial_\nu{X}_{(0)})+\nonumber\\&&X_{(1)}\partial_\mu(\sqrt{-g}g^{\mu\nu}\partial_\nu{X}_{(1)})\Bigg],\eea
 where two external ``coordinates'' are defined as
 \be\label{ct-1}
 X_{(0)}=\vh_{0}\Omega, ~~~~~~~~~~X_{(1)}=\Phi
 \ee
in accord with the standard definition of
 the general action for brane in $D/N$ dimensions
given in \cite{bn} by
 \bea S^{(D/N)}_{\mathrm{brane}}&=&-\int\!
d^Dx\!\sum_{A,B=1}^N\eta^{AB}\!\Bigg[\!\sqrt{-g}\frac{X_A
X_B}{(D-2)(D-1)}{}^{(D)}\!R(g)\!-\nonumber\\&&\!X_A\partial_\mu(\sqrt{-g}
 g^{\mu\nu} \!\partial_\nu X_B)\Bigg] \label{braneDN}
 \eea
 in the case for $D=4$,
 $N=2$ we have $\eta^{AB}=\mathrm{diag}\{1,-1\}$.
 In this case, in order to keep
 conformal invariance of the theory  (\ref{brane-m}),
   the Einstein definition of a measurable
   interval  (\ref{1-h7s}) in GR  (\ref{21-1}) should  be  replaced by
  its conformal invariant version
 as a Weyl-type ratio 
  \be \label{1-10a}
 ds_{\rm (L)}^2=\frac{ds^2}{ds_{\rm units}^2},
 \ee
 where  $ds_{\rm units}^2$ is an interval of the units
 that is defined like
 the Einstein definition of a measurable  interval  (\ref{1-h7s}) in GR.

 From the relativistic brane viewpoint
 the Einstein GR (\ref{21-1}) with the conformal coupling
 scalar field  looks like the Ptolemaeus absolutizing
 of the present-day (PD) value of one of ``coordinates'' in
 field ``superspace'' of events $[X_{(0)}|X_{(1)}]$, in our case
 it is \be\label{P-1}
 X_{(0)}\Big|_{\rm PD}=\vh_0.
 \ee
  This is equivalent to  fixation of units
 of measurements \cite{039}. Another choice of independent degrees of freedom
 in the brane theory  (\ref{brane-m}) is the unit spatial metric determinant
 \be\label{L-1}
 |g^{(3)}_{(\rm L)}|=1
 \ee
  known as the Lichnerowicz variables
 \cite{lich}. In the last case, a relativistic system in
  each frame has its proper units, like a particle in each frame in
 classical mechanics has its proper initial position and velocity (Galilei),
 and a relativistic particle in each frame in special relativity has
 its proper time (Einstein). The  ``relative units''  (\ref{1-10a})
 supposes  a new type of conformal cosmology (CC) \cite{039,bpzz,zakhy}
 with   another
 definition of the ``measurable distance'' (\ref{1-10a}),
  instead of the standard cosmology (SC)
 with absolute units (\ref{1-h7s}) \cite{f22}.
 The
  ``relative units'' (\ref{1-10a}) in CC
   excludes the expansion
 of the ``measurable'' volume of the Universe in the process of
 its cosmological evolution, as this volume does not depend on
 any scale factor including the cosmological one, whereas
 all masses in CC including the Planck one are scaled by
 cosmological scale factor. The
 relative  ``measurable distance'' (\ref{1-10a}) in CC  explains
 the SN data on the luminosity-distance -- redshift relation
 \cite{snov,SN,riess1}
 by the rigid state without $\Lambda$ -- term \cite{039,zakhy}.
Thus, a conformal-invariant relativistic brane  (\ref{brane-m})
 is a more general theory
 than Einstein GR  (\ref{21-1}), and is reduced to GR for the absolute
 units  (\ref{P-1}) or to the scalar version of the Weyl
 conformal theory in terms of the Lichnerowicz variables (\ref{L-1}).
 In the following, we call the theory (\ref{brane-m}) with condition
 (\ref{L-1}) the Conformal Relativity (CR).

 The problem is to determine
  the measurable Planck mass and cosmological scale factor
 in both the GR
 (\ref{21-1})   and CR (\ref{brane-m}). Measurable quantities
 are determined by a frame of reference to initial data in both
 the ``external superspace of events'' $[X_{(0)},X_{(1)}]$ and
 ``internal''
  Riemannian space-time $(x^0,x^k)$.

\section{Reference frame in external ``superspace of events''}

\subsection{Distortion  of GR by the conformal coupling scalar field}

 One can see that the conformal coupling scalar field $\Phi$
 distorts the Newton coupling constant in the Hilbert
  action (\ref{21-1}) distinguished by (\ref{P-1})
 \be\label{22-1}
 S_{GR+SF,\xi=1}=\int d^4x\sqrt{-g}
 \left[-\left(1\!-\!\frac{|\Phi|^2}{\vh_0^2}\right)\frac{\vh_0^2}{6}R(g)
 +g^{\mu\nu}\partial_\mu\Phi\partial_\nu\Phi
 \right],
 \ee
 This distortion
 changes
 the Einstein equations
 and their standard solutions
 of   type of Schwarzschild one and other \cite{B74,PPG,PS}
   due to  the coefficient
   $\left[1\!-\!{|\Phi|^2}/(\vh_0^2)\right]$.
   This coefficient
 restricts  region of   a scalar field  motion
  by the
 condition $|\Phi|^2\leq {\vh_0^2}$,
  because in other region $|\Phi|^2\geq {\vh_0^2}$
  the sign before the 4-dimensional curvature is changed in the
  Hilbert action  (\ref{21-1}). The rough analogy
   of this restriction is the light cone in special relativity
   which defines the physically admissible region of a particle
   motion.

  \subsection{The Bekenstein's  transformation of the Higgs field}

In order to keep the Einstein theory  (\ref{21-1}),  one needs to
consider only the field configuration such that $|\Phi|^2\leq
{\vh_0^2}$.
  For this case one can introduce  new  variables \cite{B74}
\bea\label{9-h11}
 g_{\mu\nu}&=&g_{\mu\nu}^{\rm (B)}\cosh^2 Q,
  \\\label{9-h6}
 |\Phi|^2&=&\vh_0^2\sinh^2 Q
  \eea
 considered in \cite{PPG,PS}.
 These
 variables restore the initial Einstein -- Hilbert action
 \be\label{22-2}
 S_{GR+SF,\xi=1}=\int d^4x\sqrt{-g}
 \vh_0^2\left[-\frac{R(g_{\rm (B)})}{6}+g_{\rm (B)}^{\mu\nu}
 \partial_\mu Q \partial_\nu Q\right].
 \ee
 One can see that \emph{the  Bekenstein
 transformation converts the ``conformal  coupling'' scalar field
 with the  weight $n=-1$ into
 the  ``minimal coupling''}  angle $Q$
 of the scalar -- scale mixing
 that looks like a {\it scalar graviton} with the conformal weight $n=0$.

\subsection{Choice of ``coordinates''  in  brane ``superspace of events''}

 The analogy of GR (\ref{21-1}) with  a relativistic brane (\ref{brane-m})
 distinguished by the condition (\ref{L-1})
   allows us to formulate the choice of variables (\ref{9-h11}) and (\ref{9-h6})
   as a choice of the ``frame'' in the brane ``superspace of events''
 \bea\label{br-1}
 \widetilde{X}_{(0)}&=&\sqrt{X^2_{(0)}-X^2_{(1)}}, \\
 Q&=&\rm{arc}\coth \frac{X_{(0)}}{X_{(1)}}
 \eea
As we have seen above the argument in favor of the choice of these
variables is the definition
  of
   the measurable value of the Newton constant
 \be\label{nc-1}
  G=\frac{8\pi}{3}\widetilde{X}_{(0)}^{-2}\Big|_{\rm present-day}=
  \frac{8\pi}{3}\vh^{-2}_0\ee
    as the present-day value
 of  the ``coordinate'' $\widetilde{X}_{(0)}=\vh_0$.

 In the case the action (\ref{brane-m}), (\ref{L-1})  takes the form
\bea\nonumber
&&S_{GR}[g^{\Omega}]+S_{SF,\xi=1}[g^{\Omega},\Phi^{\Omega}]=
S_{\mathrm{brane}}^{(D=4/N=2)}[X_{(0)},X_{(1)}]\!=\\\nonumber
&&\!\int\! d^4x\!\Bigg[\sqrt{-g_{(L)}}\widetilde{X}_{(0)}^2\!
\left(-\frac{{}^{(4)}\!R(g_{(L)})}{6}+g_{(L)}^{\mu\nu}\partial_\mu
Q\partial_\nu Q\right)\,+\\&&\label{brane-m3}
\widetilde{X}_{(0)}\partial_\mu\left(\sqrt{-g_{(L)}}
g_{(L)}^{\mu\nu}\partial_\nu \widetilde{X}_{(0)}\right)\Bigg].\eea
 This form is  the brane
 generalization of the relativistic conformal mechanics
\bea\label{confm-1}
S_{\mathrm{particle}}^{(D=1/N=2)}[X_{0},Q_{0}]\!&=& \!\int\!
ds\!\left[{X}_{0}^2\! \left(\frac{d
Q_{0}}{ds}\right)^2-\left(\frac{d {X}_{0}}{ds}\right)^2
\right];~~~\\\label{confm-2} &&ds = dx^0e(x^0).\eea In the following
this conformal mechanics will be considered as a simple example.

 \section{Reference frame in internal Riemannian space-time}

\subsection{The Dirac --- ADM parametrization}

  Recall that the Hamiltonian approach to GR is formulated
   in a specific frame of reference
  in terms of the Dirac -- ADM parametrization of metric \cite{dir,ADM}
  defined as
 \bea \label{1adm}
 ds^2&=&g^{\rm(B)}_{\mu\nu}dx^\mu dx^\nu~~~~~~~~~~~
 ~\equiv\omega^2_{(0)}-\omega^2_{(b)}\\\label{2adm}
 \omega_{(0)}&=&\psi^6N_{\rm d}dx^0~~~~~~~~~~~~~~
 ~\equiv \psi^2 ~\omega^{(L)}_{(0)}\\\label{3adm}
 \omega_{(b)}&=&\psi^2 {\bf e}_{(b)i}
 (dx^i+N^i dx^0)\equiv \psi^2 ~\omega^{(L)}_{(b)}
 \eea
 here triads ${\bf e_{(a)i}}$ form the spatial metrics with $\det |{\bf
 e}|=1$, $N_{\rm d}$ is the Dirac lapse function, $N^i$ is  shift
 vector and $\psi$ is a determinant of the spatial metric and
 $\omega^{(L)}_{(\mu)}$ are the Lichnerowicz simplex distinguished by
 the condition of the unit determinant (\ref{L-1}).
 In terms of these metric components the GR action takes the form
 \bea \label{6-1}
 S[\vh_0|F,Q]= \!\int\! dx^0 \vh_0^2 \!\int
 \!d^3x\Bigg[\!-\!\psi^{12}\! N_{\rm d}\frac{{}^{(4)}R(g_{})}{6}~&+&
 \frac{(\partial_0Q\!-\!N^k\partial_kQ)^2}{N_{\rm d}}
 \!-\nonumber\\&&\!N_{\rm d} \psi^8\partial_{(b)}\!Q\partial_{(b)}Q \!\Bigg],
 \eea
  where  $\partial_{(b)}Q={\bf e}^k_{(b)}\partial_k Q$ and
  ${}^{(4)}R(g_{})$ is given in  Appendix (see Eq. (\ref{Asv11})).
This action  is invariant with respect to
  transformations \cite{vlad}
 \bea \label{zel}
 x^0 &\rightarrow& \tilde x^0=\tilde x^0(x^0)\\\label{zel2}
 x_{i} &\rightarrow&  \tilde x_{i}=\tilde
 x_{i}(x^0,x_{1},x_{2},x_{3}),\\
 \label{kine}
 \tilde N_d &=& N_d \frac{dx^0}{d\tilde x^0},\\\tilde N^k&=&N^i
 \frac{\partial \tilde x^k }{\partial x_i}\frac{dx^0}{d\tilde x^0} -
 \frac{\partial \tilde x^k }{\partial x_i}
 \frac{\partial x^i}{\partial \tilde x^0}~.
 \eea
 This group of diffeomorphisms  conserves
  a family of constant-time hypersurfaces,
  and is commonly known as the {``kinemetric''} subgroup of the group of
  general coordinate  transformations $x^{\mu} \rightarrow
   \tilde x^{\mu}=\tilde
 x^{\mu}(x^0,x^{1},x^{2},x^{3})$.
    The {``kinemetric''} subgroup contains
 reparametrizations of the coordinate evolution parameter $x^0$.
 This means that in  finite space-time the coordinate evolution parameter
  $x^0$ is
 not measurable quantity, like the coordinate evolution parameter $x^0$
 in the relativistic conformal mechanics (\ref{confm-1}) that is
 invariant with respect to diffeomorphisms $x^0 \rightarrow
   \tilde x^0=\tilde
 x^0(x^0)$,
 because both  parameters $x^0$ are not diffeo-invariant.

 The relativistic mechanics  (\ref{confm-1}) has two diffeo-invariant
 measurable times.
  They are
 the geometrical interval (\ref{confm-2}) and the time-like
 variable  $X_{0}$ in the external ``superspace of events''.
 The relation between these two ``times'' $X_{0}(s)$
 are conventionally treated as  a relativistic
 transformation.
 The main problem is to point out similar
  two measurable time-like diffeo-invariant
 quantities in both GR (\ref{6-1}) and a brane  (\ref{brane-m3}).

\subsection{External diffeo-invariant evolution parameter as zero mode
 in finite volume}

 The brane/GR correspondence (\ref{brane-m})
 and special relativity (\ref{confm-1}) allows us to treat an external time
  as homogeneous component of the time-like external ``coordinate''
  $\widetilde{X}_{(0)}(x^0,x^k)$
  identifying this homogeneous component with
 the cosmological scale factor $a$
  \be\label{nc-2}
  \widetilde{X}_{(0)}(x^0,x^k)\to\vh_0a(x^0)=
  \vh(x^0)
  \ee
  because  this factor is
 introduced in the cosmological perturbation theory \cite{lif}
 by the scale transformation of the metrics (\ref{ct-1}) too
  \be\label{ct-2}
 g_{\mu\nu}=a^2(x^0){\widetilde{g}}_{\mu\nu}.
 \ee
Recall that, in this case, any field $F^{(n)}$ with the conformal
weight $(n)$ takes the form
 \be\label{F}
 F^{(n)}=a^n(x_0) {\widetilde{F}}^{(n)}. \ee
 In particular,
 the
   curvature
 $
 \sqrt{-g}\,\,{}^{(4)}R(g)=a^2\sqrt{-{\widetilde{g}}}\,\,{}^{(4)}R({\widetilde{g}})-6a
 \partial_0\left[{\partial_0a}\sqrt{-{\widetilde{g}}}~
 {\widetilde{g}}^{00}\right]$ 
  can be expressed in terms of
   the new lapse
 function ${\widetilde{N}_d}$ and spatial determinant ${\widetilde{\psi}}$ in
 Eq. (\ref{1adm})
 \be \label{lfsd}
 {\widetilde{N}}_d=[\sqrt{-{\widetilde{g}}}~{\widetilde{g}}^{00}]^{-1}=a^{2}{N}_d,~~~~~~~~
 {\widetilde{\psi}}=(\sqrt{a})^{-1}\psi.
 \ee
In order to keep the number of variables in GR, in contrast to
\cite{lif}, we identify $\log \sqrt{a}$ with
 the  spatial volume ``averaging'' of $\log{\psi}$,
 and $\log{\widetilde{\psi}}$, with the nonzero Fourier harmonics \cite{242,242a}
 \be\label{1non1}
 \log \sqrt{a}=\langle \log{\psi}\rangle\equiv\frac{1}{V_0}\int
 d^3x\log{\psi},~~~~~~
 \langle\log{\widetilde{\psi}}\rangle \equiv 0
\ee
 here
 the Lichnerowicz diffeo-invariant volume $V_0=\int d^3x$ is
 introduced. One should emphasize  that modern cosmological models
 \cite{lif}
 are considered in the finite space and ``internal finite time'' in a reference frame
 identified with the frame of the Cosmic Background Microwave Radiation.

 A scalar field can be also presented as a sum of a zero Fourier
  harmonics and nonzero ones like (\ref{1non1})
 \bea\label{z-s1}
 Q= \langle Q\rangle+\overline{Q}; ~~~~\langle\overline{Q}\rangle=0
 \eea
 After the separation of all zero modes the action  (\ref{6-1}) takes the form
\be \label{6-6}
 S[\vh_0|F,Q]= S[\vh|\widetilde{F},\overline{Q}]+
 \underbrace{V_0\!\int\! dx^0  \!
\frac{1}{{N}_0}\left[ \vh^2\left(\frac{d \langle
Q\rangle}{dx^0}\right)^2-\left(\frac{d
\vh}{dx^0}\right)^2\right]}_{zero-mode~contribution};
 \ee
 here
 \bea \label{6-4}
 S[\vh|\widetilde{F},\overline{Q}]= \!\int\! dx^0 \vh^2 \!\int
 \!d^3x\Bigg[\!-\!\widetilde{\psi}^{12}\! \widetilde{N}_{\rm d}
 \frac{{}^{(4)}R(\widetilde{g}_{})}{6}~&+&
 \frac{(\partial_0\overline{Q}\!-\!N^k\partial_k\overline{Q})^2}{\widetilde{N}_{\rm d}}
 \nonumber\\&-&\!\widetilde{N}_{\rm d} \widetilde{\psi}^8\partial_{(b)}\!\overline{Q}\partial_{(b)}\overline{Q} \!\Bigg]
 \eea
 repeats action $S[\vh_0|F,Q]$ (\ref{6-1}), where $[\vh_0|F,Q]$ are replaced by
 $[\vh|\widetilde{F},\overline{Q}]$, and
 \be \label{6-5}
 \frac{1}{N_{0}}=\frac{1}{V_{0}}\int\frac{{d^3x}}{\widetilde{N}_{d}}\equiv
 \left\langle \frac{1}{\widetilde{N}_{d}}\right\rangle
 \ee
 is the homogeneous component of the lapse function.
 The action of the local variables  (\ref{6-4}) determines
 the correspondent local energy density for the local variables
 \be \label{6-9e}-\widetilde{T}_{\rm d}=\frac{\delta
 S[\vh_0|{\widetilde{F}},{\overline{Q}}]} {\delta
 \widetilde{N}_{\rm d}}.
 \ee

\subsection{``Internal diffeo-invariant homogeneous  time''}
 The homogeneous component of the lapse function (\ref{6-5})
 $N_0$
 determines difeo-invariant local lapse function
\be \label{6-8}
 \mathcal{N}={\widetilde{N}_{\rm
 d}}{\langle{\widetilde{N}_{\rm d}^{-1}}\rangle},~~~~~
 \langle\mathcal{N}^{-1}\rangle=1,
\ee and the ``internal diffeo-invariant homogeneous time'' with its
derivative  \be \label{6-9}
 \int{dx^0}N_{0}=\zeta,~~~~f'=\frac{df}{d\zeta}.
 \ee

\subsection{Resolution of the energy constraints}

 The action principle for the $S[\vh_0|{F},{Q}]$
 with respect to the lapse function $\widetilde{N}_{\rm d}$
 gives us the energy constraints equation
 \be\label{6-7}
\frac{1}{\mathcal{N}^2}\left(\vh'^2-\vh^2 {\langle
Q\rangle'}^2\right)-\widetilde{T}_{\rm d}=0.
 \ee
 This equation is the algebraic one with respect to
 the diffeo-invariant lapse function ${\cal N}$ and has solution
 satisfying the constraint  (\ref{6-8})
 \be\label{6-10}
 {\cal N}=
 \frac{\langle\widetilde{T}_{\rm
 d}^{1/2}\rangle}{\widetilde{T}_{\rm d}^{1/2}}.
 \ee
 The substitution of this solution into the energy constraint
 (\ref{6-7}) leads to the cosmological type equation
\be\label{6-11}
 \vh'^2=\vh^2 {\langle Q\rangle'}^2+{\langle(\widetilde{T}_{\rm
 d})^{1/2} \rangle}^2\equiv \rho_{\rm tot}(\vh)=
 \frac{P_{\langle Q\rangle}^2}{4V_0^2\vh^2}+{\langle(\widetilde{T}_{\rm
 d})^{1/2} \rangle}^2
 \ee
 here
 the total energy density $\rho_{\rm tot}(\vh)$ is
 split on the sum of  the energy density of local fields
 ${\langle(\widetilde{T}_{\rm
 d})^{1/2} \rangle}^2$ and the zero mode one, where
 \be\label{6-12}
 P_{\langle Q\rangle}=2V^0{\vh^2\langle Q\rangle'}\equiv 2V^0p_0
 \ee
 is the scalar field zero mode momentum that is
an integral of motion of the considered model because
 the action does not depend on $\langle Q\rangle$.
  The value
 of the local energy density onto
  solutions of  motion equations  depends on only $\vh$ too,
 because momentum of the external time $\vh$
 \be\label{ecs} P_\vh=2V_0\vh'= \pm 2V_0\sqrt{\frac{p_0^2}{\vh^2} +
 {\langle(\widetilde{T}_{\rm
 d})^{1/2} \rangle}^2}\equiv\mp E_\vh\ee
   can be considered as the Hamiltonian of evolution
   in the ``superspace of events''.
 The value of  the momentum $P_\vh=\pm E_\vh$ onto
  solutions of motion equation
 is defined as an energy of the universe, in accord with the
 second N\"other theorem removing momenta by constraints following
 from diffeomorphisms. We can see that the dimension of the
 group of diffeomorphisms (\ref{zel}) and (\ref{zel2})
 coincides with the dimension of the first class constraint
 momenta, because the local part of the energy constraint   (\ref{6-7})
 determines the diffeo-invariant local lapse function ${\cal N}$   (\ref{6-10}).

 The solution  (\ref{ecs}) gives the diffeo-invariant constraint-shell action
 (\ref{100}) obtained in Appendix
  \be\label{ecs2}
 S_{{\cal H}=0}\!=
\int\limits_{\vh_I}^{\vh_0}d\widetilde{\vh}
\left\{\int\limits_{V_0}^{
 } d^3x\sum\limits_{\widetilde{F} }
 P_{\widetilde{F}}\partial_\vh \widetilde{F}
 \mp2E_\vh\right\}.
\ee
  The GR version of the Friedmann equation (\ref{6-11}) leads to the
  diffeo-invariant Hubble law
  as the relation between the geometric time  (\ref{6-9})
  and the cosmological scale factor $\vh=\vh_0 a$ \cite{242,242a}
 $$\zeta_{(\pm)}=\int dx^0N_0=\pm\int^{\vh_0}_{\vh_I}
{d\vh}~\left[{p_0^2}/{\vh^2} +
 {\langle(\widetilde{T}_{\rm
 d})^{1/2} \rangle}^2\right]^{-1/2}\geq 0,$$ where
 \be\label{1-37} \widetilde{T}_{\rm d}= \frac{4\vh^2}{3}{\widetilde{\psi}}^{7} \triangle
{\widetilde{\psi}}+
  \sum\limits_{I} \vh^{I/2-2}{\widetilde{\psi}}^{I}\overline{{\cal T}}_I; \ee
 here $\overline{\cal{T}}_I$ is partial energy density
  marked by the index $I$ running, in general case, a set of values
   I=0 (stiff), 4 (radiation), 6 (mass), 8 (curvature)\footnote{$\Lambda$-term
    corresponds to $I=12$}
 in correspondence with a type of matter field
 contributions (see Appendix, Eqs. (\ref{h32}) and (\ref{h35})).

  The second class Dirac condition
 of the minimal 3-dimensional hyper-surface [5]
\be\label{11-42}
 p_{{\widetilde{\psi}}}=0 \to
 (\partial_\zeta-N_{(b)}\partial_{(b)})\log{
 {\widetilde{\psi}}}=\frac16\partial_{(b)}N_{(b)},
 \ee
 is included in order to give a positive value of the Hamiltonian density
 $\widetilde{T}_{\rm d}$  given by Eq. (\ref{1-37}) and
 Eq. (\ref{h32}) in the Appendix. The equations  (\ref{6-10})
  and
 $\widetilde{T}_{\psi}-\langle\widetilde{T}_{\psi}\rangle=0$
 (where $\widetilde{T}_{\psi}=T_{\psi}[\vh|\widetilde{\psi}]$
 and $T_{\psi}$ is given by Eq.  (\ref{1-37ab}))
 determine  the lapse function ${\cal N}$ and the scalar component
 $\widetilde{\psi}$.

 Thus, we give the diffeo-invariant formulation of the GR with
 the conformal scalar field in the comoving CMB frame compatible with
  the Einstein equations $T_{\rm d}=T_{\rm \psi}=0$
 and their Schwarzschild-type solutions $\triangle \psi=0$,
  $\triangle (\psi^7 {\cal N})=0$ in the infinite volume limit
 \cite{242},
  in contrast to all other
 approach to  a scalar field  in GR \cite{linde,MFB}.

 The
 special relativity identification of the brane external time-like
 ``coordinate''
 with the diffeo-invariant evolution parameter (\ref{nc-2})
 $$\widetilde{X}_{(0)}(x^0,x^k)=
 \sqrt{X^2_{(0)}-X^2_{(1)}}=\vh(\zeta)\widetilde{\psi}^2$$
 arising after resolving energy constraint with respect
 its momentum $P_\vh$  is in agreement with the Hamiltonian version
 \cite{242,242a,bpzz} of cosmological
 perturbation theory \cite{lif} identifying this external time-like
 ``coordinate'' with
  the cosmological scale factor $a(\eta)=\vh(\eta)/\vh_0$
 provided that
\begin{enumerate}
 \item the cosmological scale factor
 is a zero mode of the scalar
 component of metric (but it is not additional variable
 as it is supposed in the accepted  cosmological perturbation theory \cite{lif})

 \item the conformal time in
 the  redshift -- luminosity
 distance relation is gauge-invariant measurable quantity,
 in accord with
 the Dirac definition of observable quantities as diffeo-invariants
 (but it is not diffeo-variant quantity
 as an object of Bardeen's gauge transformations
 in the accepted perturbation theory \cite{lif}),

 \item the initial
 datum $a(\eta=0)=a_I$ is free from the current data
$a'(\eta_0)$ and fundamental parameter $M_{\rm Planck}$ of the
motion equations (because   the Planck epoch data
$a(\eta=0)=a'(\eta_0)/M_{\rm Planck}$  violate the causality
principle in the constraint action (\ref{ecs2}) and they are the
origin of numerous problems in the Inflationary Model \cite{linde}),

 \item there is the vacuum as a state with the minimal energy
 as explicit resolving the energy constraint. The vacuum postulate
 is provided by the  second class Dirac condition
 of the minimal 3-dimensional hyper-surface [5]
(\ref{11-42}) that removes the kinetic perturbations of the
 accepted cosmological perturbation theory explaining the power
 CMB spectrum \cite{MFB}.
\end{enumerate}
 However, the
 accepted cosmological perturbation theory \cite{MFB} omitted the potential
 perturbations  going  from the scalar metric component
 $\widetilde{\psi}=1-\Psi/2$
 in partial energy density
 (\ref{1-37})
that leads to additional fluctuations of the CMB temperature
\cite{242}.

\section{Cosmology and the Cauchy problem of the zero mode dynamics}
 The conformal-invariant unified theory alternative 
  in
 the homogeneous approximation
 \bea\label{2.3-1}
 w(x^0,x^k)&=&\vh(x^0)\equiv\vh_0a(x^0),\\\label{2.3-2}
 Q(x^0,x^k)&=& \langle Q\rangle (x^0)\equiv\frac{1}{V_0}\int d^3x
 Q(x^0,x^k),\\\label{2.3-3}
 N_d(x^0,x^k) &=& N_0(x^0), \\\label{2.3-4}
 N_0(x^0)dx^0 &=& d\eta
 \eea
 leads to
 the
 cosmological model given by the action
 \bea\label{2.3-6} S&=&V_0\int dx^0
\left[\frac{-(\partial_0\vh)^2+\vh^2(\partial_0\langle
Q\rangle)^2}{N_0}
\right]=\\
\label{2.3-7} &=&\int dx^0 \left\{P_Q\frac{d}{dx^0} \langle Q\rangle
-P_\vh\frac{d}{dx^0} \vh +\frac{N_0}{4V_0} \left[P_\vh^2
-\frac{P^2_Q}{\vh^2}\right] \right\}\eea
 where $V_0=\int d^3x$ is finite coordinate volume,
 \bea\label{2.3-9}
 P_\vh&=&2V_0\vh'\equiv2V_0\frac{d\vh}{d\eta},\\\label{2.3-10}
 P_Q&=&2V_0\,\vh^2\,{\langle Q\rangle}'
 \eea
 are canonical conjugated momenta, $\eta = \int dx^0 N_0(x^0)$
  is the conformal time.

 The energy constraint in the model
 \bea\label{2.3-11}
 P_\vh^2-E_\vh^2=0;~~~~~~~~~~~~E_\vh=\frac{|P_Q|}{\vh}
\eea repeat completely the cosmological equations
 of the GR in the case of
 a rigid equation of state $\Omega_{\rm rigid}=1$
\bea\label{2.3-12}
 \vh_0^2 a'^2=\frac{P_Q^2}{4V_0^2\vh^2}\equiv \frac{\rho_0}{a^2}
 =H_0^2 \frac{\Omega_{\rm rigid}}{a^2},
\eea
  where $P_Q$ is the constant of the motion, because
    \be\label{2.3-14}
  P_Q'=0.
  \ee
 The solution of these equations take the form
\bea\label{2.3-16}
 \vh(\eta)=\vh_I\sqrt{1+2{\cal H}_I\eta}, ~~~~
 Q(\eta)=Q_I+\log {\sqrt{1+2{\cal H}_I\eta}},
\eea
 where
\bea\label{2.3-17}
 \vh_I&=&\vh(\eta=0),\\\label{2.3-18}
  Q_I&=&Q(\eta=0),~~~~P_Q={\rm const}
\eea
 are the ordinary initial data. These data do not depend on
 the current values of variables $\vh_0=\vh_0a(\eta=\eta_0)$
 in contrast to the Planck epoch one, where the initial data
 of the scale variable
$a_I=a(\eta=0)=a'(\eta_0)/M_{\rm Planck}$  are determined by
 its velocity at present-day epoch. We have seen above that this determination
    violates the causality principle in the constraint-shell action
(\ref{ecs2}).

\section{Conclusion}

We convince that the conformal symmetry is the way for
classification of scalar field dynamics in GR. Conformal
transformation allows us to convert the conformal coupling scalar
field into a conformal relativistic brane without any dimensional
parameter. Spontaneous conformal symmetry breaking in this case can
be provided by initial data.

Consideration  of the diffeo-invariant initial data in a specific
frame differs our approach to scalar field in this paper from other
approaches to this problem. A definition of initial data as
diffeo-invariant measurable quantities supposes two distinguished
reference frames - the observer rest frame and the observable
comoving frame. In particular, comoving frame of the Universe is
identified with the CMB frame that differs from the rest frame by
the non zero dipole component of the temperature fluctuations. {\it
Differences} between these two frames lie in essence of all
principles of relativity including the Galilei's relativity as a
{\it difference} of initial positions and velocities, the Einstein's
relativity as a {\it difference} of proper times, and the Weyl's
relativity of a {\it difference} of units. A  definition of
reference frame, in our paper, is based on the Fock simplex (in
order to separate diffeomorphism from frame transformation), the
Dirac--ADM parametrization of metric (in order to classify of the
metric components), and the Zel'manov ``kinemetric'' diffeomorphisms
as parametrizations of the internal coordinate (in order to identify
the diffeo-invariant
  evolution parameter with the cosmological scale factor as
the zero mode of metric determinant and to define the energy as the
constraint-shell value of the scale momentum).
 Finally, the Hamiltonian
action in GR coincides with the relativistic brane one,
 where time-like external
 coordinate plays the role of the diffeo-invariant
 evolution parameter in the field ``superspace of events'',
 and its momentum plays the role of the energy
 in accord with special relativity given in the Minkowskian
 space of events.
  Therefore, the generalization of the Dirac Hamiltonian approach to
  the conformal coupling scalar field gives us the
possibility to restore the universal Hamiltonian description of
relativistic brane-like systems with the action $S^{D/N}$ with any
number of external and internal coordinates. Thus, we show that the
Dirac Hamiltonian approach to the conformal coupling scalar field in
GR coincides with the similar consideration of the conformal brane
$S^{D=4/N=2}$. Both these theories (the conformal coupling scalar
field and the brane) lead to the rigid state in agreement with the
SN data on the luminosity-distance -- redshift relation
\cite{snov,SN,riess1}  in framework of the conformal cosmology
\cite{039,bpzz,zakhy}, where the Weyl relativity of units
(\ref{1-10a}) is supposed.

\section*{Acknowledgements}

All authors are grateful to Dmitry Kazakov and Igor Tkachev for
discussions of statement of the problem of the rigid state in the
modern cosmology. The authors are grateful to B.M. Barbashov, K.A.
Bronnikov, V.V. Kassandrov, E.A. Kuraev, D.G. Pavlov, Yu.P. Rybakov
and A.F. Zakharov for interesting and critical discussions. \L.A.
Glinka is thankful to the Bogoliubov-Infeld program of grants for
partial financial support. R.P. Kostecki is thankful to \L.A.G. and
V.N.P. for hospitability.

\section*{The Appendix. The Dirac--ADM approach to GR} The
 Hilbert action $S=S_{\rm GR}+S_Q$
  in terms of the Dirac -- ADM variables
  (\ref{2adm}) and (\ref{3adm})
 is as follows \bea\nonumber \label{Asv11}
 &&S_{\rm GR}= -\int d^4x\sqrt{-g}\frac{\vh_0^2}{6}~{}^{(4)}R(g)=\int d^4x
 ({\mathcal{K}}[\vh_0|
 {g}]-{\mathcal{P}}[\vh_0|{g}]+{\mathcal{S}}[\vh_0|{g}])\nonumber\\
 &&S_{\rm Q}
 =\vh_0^2\int dx^0 d^3x \Bigg[\!\!
 \frac{(\partial_0Q\!-\!N^k\partial_kQ)^2}{N_{\it d}}
 -N_{\it d} \psi^8(\partial_{(b)}Q)^2 \Bigg],
 \eea
 where
\bea
 {\mathcal{K}}[\vh_0|e]&=&{{N}_d}\vh_0^2\left(-{\vphantom{\int}}4
 {  {v}}^2+\frac{v^2_{(ab)}}{6}\right),
 \label{k1}\\
 {\mathcal{P}}[\vh_0|e]&=&\frac{{N_d}\varphi_0^2{\psi}^{7}}{6}\left(
 {}^{(3)}R({\bf e}){\psi}+
 {8}\triangle{\psi}\right),
 \label{p1}\\
 {\cal S}[\vh_0|e]&=&2\varphi_0^2\left[\partial_0{v_{\psi}}-
 \partial_l(N^l{v_{\psi}})\right]-\frac{\varphi^2_0}3 \partial_j[\psi^2\partial^j (\psi^6
 N_d)]\label{0-s1};
 \eea
 are the kinetic and  potential terms,
 respectively,
  \bea\label{proi1}
 {v}&=&\frac{1}{{N_d}}\left[
 (\partial_0-N^l\partial_l)\log{
 {\psi}}-\frac16\partial_lN^l\right],\\
 v_{(ab)}&=&\frac{1}{2}\left({\bf e}_{(a)i}v^i_{(b)}+{\bf
 e}_{(b)i}v^i_{(a)}\right),\\\label{proizvod}
 v_{(a)i}&=&
 \frac{1}{{N_d}}\left[(\partial_0-N^l\partial_l){\bf e}_{(a)i}
+ \frac13 {\bf
 e}_{(a)i}\partial_lN^l-{\bf e}_{(a)l}\partial_iN^l\right],\\\label{proizvod-p}
 v_{Q}&=&\frac{\partial_0Q-N^k\partial_kQ}{N_{\it d}}
 \eea
 are velocities of the metric components,
   ${\triangle}\psi=\partial_i({\bf e}^i_{(a)}{\bf
 e}^j_{(a)}\partial_j\psi)$ is the covariant Beltrami--Laplace operator,
 ${}^{(3)}R({\bf{e}})$ is a three-dimensional curvature
 expressed in terms of triads
   ${\bf e}_{(a)i}$:
\be \label{1-17}
 {}^{(3)}R({\bf e})=-2\partial^{\phantom{f}}_{i}
 [{\bf e}_{(b)}^{i}\sigma_{{(c)|(b)(c)}}]-
 \sigma_{(c)|(b)(c)}\sigma_{(a)|(b)(a)}+
 \sigma_{(c)|(d)(f)}^{\phantom{(f)}}\sigma^{\phantom{(f)}}_{(f)|(d)(c)}.
 \ee
 Here
 \be\label{1-18} \sigma_{(a)|(b)(c)}=
 {\bf e}_{(c)}^{j}
 \nabla_{i}{\bf e}_{(a) k}{\bf e}_{(b)}^{\phantom{r}k}=
 \frac{1}{2}{\bf e}_{(a)j}\left[\partial_{(b)}{\bf e}^j_{(c)}
 -\partial_{(c)}{\bf e}^j_{(b)}\right]
  \ee
  are the coefficients of the spin-connection (see \cite{242a}),
  \be\nabla_{i}{\bf e}_{(a) j}=\partial_{i}{\bf e}_{(a)j}
  -\Gamma^k_{ij}{\bf e}_{(a) k}\ee are covariant derivatives, and
  $\Gamma^k_{ij}=\frac{1}{2}{\bf e}^k_{(b)}(\partial_i{\bf e}_{(b)j}
  +\partial_j{\bf e}_{(b)i})$. The  canonical conjugated momenta are
\bea \label{1-32}{p_{\psi}}&=&\frac{\partial {\cal
K}[\vh_0|e]}{\partial
 (\partial_0\ln{{{\psi}}})}~=-8\vh_0^2{{v}},
 \\\label{1-33}
 p^i_{(b)}&=&\frac{\partial {\cal K}[\vh_0|e]
 }{\partial(\partial_0{\bf e}_{(a)i})}
 =\frac{\vh^2}{3}{\bf e}^i_{(a)} v_{(a b)},\\\label{1-34}
 P_{Q}&=&2\vh^2_0\frac{\partial_0Q-N^k\partial_kQ}{N_{\it d}}.
 \eea
The Hamiltonian action takes the form \cite{242,242a}
 \be\label{1-16} S=\int d^4x
 \left[\sum\limits_{{F=e,\log\psi,Q}
 } P_{F}\partial_0F
 -{\cal H}\right] \ee
 where
 \be\label{1-17c}
{\cal H}={N_d} T_{\rm d}+N_{(b)}
  {T}^0_{(b)} +\lambda_0{p_\psi}+
  \lambda_{(a)}\partial_k{\bf e}^k_{(a)}
 \ee
is the sum of constraints with  the Lagrangian multipliers ${N_d}$,
$N_{(b)}={\bf e}_{k(b)}N^k$, $\lambda_0$ $,\lambda_{(a)}$, and ,
$T^0_{\rm (a)}=
 -\!\!{p_{\psi}}\partial_{(a)}
 {\psi}\!+\!\frac{1}{6}\partial_{(a)}
 ({p_{\psi}}{\psi})\! +\!
 2p_{(b)(c)}\gamma_{(b)|(a)(c)}\!-\!\partial_{(b)}p_{(b)(a)}\!+\!
 P_Q\partial_{(a)}Q $
 are the components of the total
energy-momentum tensor ${T}^0_{(a)}=-\frac{\delta S}{\delta
N_{k}}{\bf e}_{k(a)}$,
 and
\bea\label{1-37a} T_{\rm d}[\vh_0|\psi]&=&-\frac{\delta S}{\delta
N_{\rm d}} = \frac{4\vh_0^2}{3}{\psi}^{7} \triangle {\psi}+
  \sum\limits_{I} {\psi}^I{\cal T}_I,\\\label{1-37ab}
   T_{\psi}[\vh_0|\psi]&=&-\psi\frac{\delta S}{\delta \psi}\equiv
  \frac{4\varphi_0^2}{3}\left[7N_d{\psi}^{7}\!
  \triangle
{\psi}+{\psi}\! \triangle \!\![N_d{\psi}^{7}]\!\right]\!+\!
 \! N_d\sum\limits_{I 
 }I {\psi}^I{\cal T}_I=0;\eea
 here $\cal{T}_I$ is partial energy density
  marked by the index $I$ running, in general case, a set of values
   I=0 (stiff), 4 (radiation), 6 (mass), 8 (curvature) 
 in correspondence with a type of matter field contributions
  \bea\label{h31}
 {\psi}^{7} \triangle
 {\psi}&\equiv&{\psi}^{7}
 \partial_{(b)}\partial_{(b)}{\psi}\\\label{h32}
 {\cal T}_{I=0}&=&\frac{6{p}_{(ab)}{p}_{(ab)}}{\vh_0^2}
 -\frac{16}{\vh_0^2}{p_{\psi}}^2+\frac{P_{Q}^2}{4\vh_0^2}
\\\label{h35}
 {\cal T}_{I=8}&=&\varphi_0^2\left[\frac{1}
  {6}R^{(3)}({\bf e})+{\partial_{(b)}Q\partial_{(b)}Q}\right],
\eea
  here ${p}_{(ab)}=\frac{1}{2}({\bf e}^i_{(a)}\widetilde{p}_{(b)i}+
  {\bf e}^i_{(b)}{p}_{(a)i})$,
 we include the Dirac local condition
 of the minimal 3-dimensional hyper-surface [5] too
\be\label{1-42}
 p_{{\widetilde{\psi}}}=0 \to
 (\partial_0-N^l\partial_l)\log{
 {\widetilde{\psi}}}=\frac16\partial_lN^l,
 \ee
 in order to obtain a positive value of the Hamiltonian density
 (\ref{h32}) after the separation of the cosmological scale factor (\ref{1non1}).

 The constraint-shell action (\ref{1-16}) after the separation of
 the zero modes (\ref{1non1}) and (\ref{z-s1}) takes the form
\bea\nonumber S\big|_{{\cal H}=0}\!\!\!&=&\!\!
 \int\! dx^0\!\int d^3x
 \sum\limits_{{F}={\psi},e,\,{Q}}P_{{F}}\partial_0F|_{\vh_0
 a=\vh}=\\\nonumber
 &=& \int dx^0
 \left\{\int\limits_{V_0}
d^3x\sum\limits_{\widetilde{F}=\widetilde{\psi},e,\, \overline{Q}}
 P_{\widetilde{F}}\partial_0 \widetilde{F}
 -P_\vh\frac{d\vh}{dx^0}+P_{\langle Q\rangle}\frac{d\langle
 Q\rangle}{dx^0}\right\}
=\\\label{100}
&=&\int\limits_{\vh_I}^{\vh_0}d{\vh}\left\{\int\limits_{V_0}
d^3x\sum\limits_{\widetilde{F}=\widetilde{\psi},e,\, \overline{Q} }
 P_{\widetilde{F}}\partial_\vh \widetilde{F}
 +P_{\langle Q\rangle}\frac{d\langle
 Q\rangle}{d\vh}
 -P_\vh\right\}.
 \eea
where $P_\vh=\pm E_\vh$ is the constraint-shell Hamiltonian in the
``superspace of events'' given by the resolving the energy
constraint (\ref{ecs}), where $\widetilde{T}_{\rm d}=T_{\rm
d}[\vh|\widetilde{\psi}]$, and $T_{\rm d}[\vh|\widetilde{\psi}]$ is
given by Eqs. (\ref{1-37a}), (\ref{h32}) and (\ref{h35}) where
$[\vh_0|{\psi}]$ is replaced by $[\vh|\widetilde{\psi}]$.

\end{document}